\documentclass[useAMS,usenatbib]{mn2e}
 
\usepackage{amssymb,amsmath}
\usepackage{graphicx} 
\usepackage{appendix}
\usepackage{color,soul}
\soulregister\citealt7
\soulregister\citet7
\soulregister\citep7
\soulregister\ref7

\usepackage[T1]{fontenc} 
\usepackage{aecompl}

\title[Radio weak lensing shear measurement]{Radio Weak Lensing Shear Measurement in the Visibility Domain - I. Methodology}
\author[Rivi et al.]{M. Rivi$^{1,2}$\thanks{E-mail: m.rivi@ucl.ac.uk}, L. Miller$^{2}$, S. Makhathini$^{3,4}$ and F.B. Abdalla$^{1,3}$\\
$^{1}$Department of Physics and Astronomy, University College London, Gower Street, London, WC1E 6BT, UK\\
$^{2}$Astrophysics, Department of Physics, University of Oxford, Keble Road, Oxford, OX1 3RH, UK\\
$^{3}$Department of Physics and Electronics, Rhodes University, PO Box 94, Grahamstown, 6140, South Africa\\
$^{4}$SKA South Africa}

\begin{document}

\maketitle

\begin{abstract}
The high sensitivity of the new generation of radio telescopes such as
the Square Kilometre Array (SKA) will allow cosmological weak lensing
measurements at radio wavelengths that are competitive with optical
surveys.  We present an adaptation to radio data of \textit{lens}fit,
a method for galaxy shape measurement originally developed and used
for optical weak lensing surveys. This likelihood method uses an
analytical galaxy model and makes a Bayesian marginalisation of the
likelihood over uninteresting parameters. It has the feature of
working directly in the visibility domain, which is the natural
approach to adopt with radio interferometer data, avoiding systematics
introduced by the imaging process. As a proof of concept, we provide
results for visibility simulations of individual galaxies with flux
density $S \ge 10\mu$Jy at the phase centre of the proposed SKA1-MID
baseline configuration, adopting 12 frequency channels in the band 950
- 1190~MHz.  Weak lensing shear measurements from a population of
galaxies with realistic flux and scalelength distributions are
obtained after natural gridding of the raw visibilities.  Shear
measurements are expected to be affected by `noise bias': we estimate
the bias in the method as a function of signal-to-noise ratio (\textrm{SNR}).
We obtain additive and multiplicative bias values that are comparable
to SKA1 requirements for $\textrm{SNR}> 18$ and $\textrm{SNR}>30$, respectively.
The multiplicative bias for \textrm{SNR}~$>10$ is comparable to that found in
ground-based optical surveys such as CFHTLenS, and we anticipate that
similar shear measurement calibration strategies to those used for
optical surveys may be used to good effect in the analysis of SKA
radio interferometer data.
\end{abstract}


\section{Introduction}
\label{sec:intro}
Weak gravitational lensing is the coherent deformation in the apparent shapes of galaxies due to the deflection of light rays by large-scale foreground matter distributions (see~\citealt{Kil15} for an overview). The measure of this distortion on cosmological scales 
is a powerful technique for estimating the total mass distribution and the relationship between the distributions of dark and baryonic matter. Its combination with redshift measurements can provide cosmological constraints on the density of dark matter and, through the growth of large-scale structure,
also on the dark energy component of the universe. Combination with other cosmological measurements may allow
tests for modifications of General Relativity.

Observationally, this field has been served so far by optical surveys since its initial detection
(\citealt{BRE00}; \citealt{WTKDB00}; \citealt{KWL00}; \citealt{VME00}), owing to the larger number densities of faint galaxies achieved in such surveys. Moreover, the redshift distribution of faint radio-detected galaxies is not known accurately, making the interpretation of the measurement very challenging.  
However, the new generation of radio telescopes, such as the Square Kilometre Array (SKA)\footnote{https://www.skatelescope.org/}, are expected to reach sufficient sensitivity to resolve radio emission of ordinary galaxies and therefore provide a large number density. For example, SKA will reach number densities of up to $\sim 3~\textrm{galaxies/arcmin}^2$ in Phase~1 and $\sim 10~\textrm{galaxies/arcmin}^2$ in Phase~2~(\citealt{Brown15}). 
This will lead weak lensing to become one of the primary science drivers in radio surveys too, with the advantage that they will access the largest scales in the Universe going beyond optical surveys, such as \textrm{LSST}\footnote{http://www.lsst.org/} and Euclid\footnote{http://www.euclid-ec.org/}, in terms of redshifts that are probed. 
Source redshifts will be available, although not at high redshifts, from HI 21 cm line observations~(\citealt{Blake04}; \citealt{YBS15}; \citealt{ABC15}), and photometric redshifts should be available from cross-correlation with faint multiband optical surveys such as LSST.
Furthermore, the radio waveband may offer unique approaches that are not available to optical surveys and 
may be used to reduce or mitigate some of the systematic effects encountered in weak lensing cosmology. 
First, the Point Spread Function (PSF) of a radio interferometer, uncertainty in which is one of the biggest causes of systematic errors in ground-based optical surveys, is largely determined by the known placement of antennas,  
and PSF variations at mid-frequencies caused by ionospheric/tropospheric phase perturbations
are expected to be smaller than PSF variations in ground-based optical surveys~(\citealt{JB2015}; \citealt{BHCB16}).
Second, it may be possible to use polarized emission as an estimator of the intrinsic (unlensed) galaxy orientation, 
although the limiting flux density, and hence galaxy number density, would be compromised in such an analysis.
Gravitational lensing does not change the position angle of the polarization emission of a galaxy and polarization is correlated with the disk structure of the galaxy~(\citealt{BB11}; \citealt{WBB15}). This technique may be used to effectively measure or correct for intrinsic galaxy alignments~\citep{JCK15}, which is likely to be one of the main astrophysical biases of weak lensing measurements, and may potentially reduce shear measurement systematics. 
HI rotational velocity measurements may also be used to reduce the impact of shape noise and intrinsic alignments. The idea, suggested by \citet{Blain02} and \citet{Morales06}, is to measure the departure from perpendicularity of the rotation axis of a disk galaxy to the major axis of the galaxy disk image and use this measure as an estimate of the shear field at the galaxy's position (see also \citealt{HKEGS13}).
Finally, by cross-correlating the shear estimators of optical and radio surveys, uncorrelated 
systematic errors may be mitigated~(\citealt{Patel10}; \citealt{DB16}; \citealt{Harrison16}; \citealt{Camera16}). 
For a general overview of radio weak lensing see~\citet{Brown15}.

However, the large field of view together with the new sensitivity regime of instruments such as SKA
will need a more detailed treatment of the systematics of radio observations. New analysis techniques and algorithm development may be required, in particular the development of highly accurate shape or shear estimation techniques suitable for a radio interferometry dataset.
Initial steps are being taken by a number of SKA pathfinders and precursor telescopes. For example the UK e-MERLIN legacy projects e-MERGE\footnote{http://www.e-merlin.ac.uk/legacy/projects/emerge.html} and SuperCLASS\footnote{http://www.e-merlin.ac.uk/legacy/projects/superclass.html} will act as training experiments for algorithms on long-baseline high-resolution observations. Other projects are planned with the upgraded JVLA interferometer: the VLASS\footnote{https://science.nrao.edu/science/surveys/vlass} survey~\citep{Brown13} and the \textrm{CHILES}\footnote{http://www.mpia-hd.mpg.de/homes/kreckel/CHILES/index.html} continuum and HI surveys. Large scale surveys with the LOFAR\footnote{http://www.lofar.org} telescope and with the SKA pathfinder telescopes, MeerKAT\footnote{http://www.ska.ac.za/meerkat/} and ASKAP\footnote{http://www.atnf.csiro.au/projects/askap} will also offer interesting opportunities for radio weak lensing studies in the run-up to Phase 1 of the SKA.

Currently, most of the techniques available for the measurement of galaxy shapes are based on measurement of galaxy images, as they were developed for optical surveys. The state-of-the-art in optical lensing measurements fits model surface brightness distributions to galaxies and combine these measurements to form an estimate of the cosmic shear. For a summary, see \citet{mandelbaum15}.
Radio interferometers do not provide directly images of the observed sky, they measure visibility data instead, that basically are the Fourier transform of the sky image at sampled points in the Fourier (uv) domain.
Such points correspond to the projected baseline formed between two antennas on the plane orthogonal to the antennas pointing direction (usually also the adopted \textit{phase centre}), and the locus traced by them during an observation as the Earth rotates yields the uv data whose Fourier transform is the PSF. The standard procedure adopted for turning visibility data into images via PSF deconvolution, such as CLEAN (\citealt{clean74}; \citealt{clean78}), is a non-linear process and the noise in radio images is highly correlated. Weak lensing has stringent requirements on image fidelity because source ellipticities must be measured accurately in order for errors on cosmological parameters to be dominated by statistics, rather than systematics.  
An investigation using images simulated through a radio pipeline presented in~\citet{Patel13}, for e-MERLIN and LOFAR, and used in~\citet{Patel15} for SKA1, shows that current iterative deconvolution methods produce images with structures in the residuals that dominate the cosmological signal, producing an analysis-induced bias far from what is required.

A more natural approach for radio weak lensing is to measure source shapes directly in the visibility domain, avoiding image reconstruction and reducing original data manipulation. This would also benefit from the fact that the noise originates  
in this domain.   
Such methods should take into account model accuracy, the fact that sources are no longer localised in the Fourier  domain and their flux is mixed together in a complicated way, which may require joint fitting. Moreover the computational challenges for a telescope such as SKA are great because the number of sources in the primary beam and the number of visibilities are very large. 
Available tools for model fitting in the visibility domain are very generic and based on models obtained from images built on the combination of basic shapes or brightness profiles \citep{MVidal14}. To optimise computational performance and model accuracy, models should be defined directly in the visibility domain, avoiding Fourier Transform operations. For a list of analytical models available in the Fourier space see Table~3 in \citet{Rowe15}, although numerically-defined models may also be interpolated to create appropriate Fourier-space models without requiring an analytic expression.

At present, the only radio weak lensing studies in the visibility domain have used shapelets~(\citealt{Refregier03}; \citealt{RB2003}), where galaxy shapes are decomposed through an orthonormal basis of functions corresponding to perturbations around a circular gaussian.
The first study,~\citet{Chang04}, used data from the Faint Images of the Radio Sky at Twenty cm (FIRST) survey~\citep{Becker95} conducted with the VLA. Shapes of radio galaxies were obtained directly from the visibilities~\citep{CR02} as shapelets are invariant under Fourier transform (up to a rescaling). A treatment of systematics that may affect the radio lensing shear estimates was also included, allowing a $3.6\sigma$ detection of cosmic shear.
The second more recent work is presented in~\citet{Patel15}, where an initial analysis of the performance of visibility plane shapelets is provided from SKA1 simulations.
Conceptual concerns about shapelets method were already raised in the optical surveys because such models do not represent galaxy shapes in a realistic way, that takes into account knowledge of galaxy structure. Moreover profiles that are not well matched to the size of the gaussian require very high order shapelet terms, such that the decomposition is unfeasible in practice~(\citealt{Melchior10}; \citealt{mandelbaum15}). 
The radio source population in the next generation of deep radio continuum surveys is expected to be dominated by late-type normal and star-forming galaxies. Their radio emission, in the range of frequencies for weak lensing observations ($\sim 1$~GHz), is dominated by the synchrotron radiation from relativistic electrons accelerated in supernova remnants~(\citealt{Condon92}; \citealt{Richards00}; \citealt{Jarvis15}), i.e. it is produced by the interstellar medium in the disk alone. A reasonable assumption is that the radio-emitting plasma has an exponential disc structure similar to that which describes the distribution of
stars in galaxy disks. The possible amplitude of bias arising from imperfect models has already been discussed in the optical case by \citet{VB10} and \citet{Miller13}.

In this paper we present an adaptation to the visibility domain of~\textit{lens}fit, a model fitting approach developed by~\citet{Miller07} and \citet{Kitching08} for optical surveys, recently used to measure galaxy shapes \citep{Miller13} in the CFHTLensS \citep{Heymans12}, CS82 (Erben et al., in preparation), KiDS \citep{Kuijken15, Hildebrandt16b} and RCSLenS \citep{Hildebrandt16a} surveys. That analysis assumed galaxies to comprise two components, disk and bulge, whereas in radio observations we are primarily interested only in a disk-like component, as discussed above.
Accordingly, in our adaption to the radio band, the method models galaxy shapes using an exponential profile (S\'{e}rsic index $n=1$) and applies a Bayesian marginalisation of the likelihood over uninteresting  parameters. We directly estimate the likelihood from the visibility data and define the model visibilities analytically. 
As a proof of concept, we present the method for the shape measurement of a single galaxy at the phase centre (Sect.~\ref{sec:radiolensfit}).   
We show results obtained from simulations of visibilities generated by using the SKA1-MID baseline configuration (details are provided in Sect.~\ref{sec:ska-simulations}). In particular in Sect.~\ref{sec:shear} we provide an estimate of the shear bias in the method. 
In Sect.~\ref{sec:multiple} we finally discuss two possible approaches to the fitting of many sources in the field of view.

\section{Weak Lensing shear estimation}
The weak lensing signal is carried by the faintest star forming (SF) galaxies. The radiation emitted by such galaxies is deflected in presence of a gravitational potential on the path to an observer. The deflection angle is approximated to first order by the Jacobian matrix $\mathbf{A}_\gamma$ of the mapping between the source and the observer,
\begin{equation}
\label{eq:transform}
\mathbf{A}_\gamma = \begin{pmatrix} 1-\kappa-\gamma_1 & -\gamma_2 \\ -\gamma_2 & 1-\kappa+\gamma_1 \end{pmatrix}
\end{equation}  
where $\kappa$ is the \textit{convergence} and the change in observed size of the source, and 
where $\mathbf{\gamma} = \gamma_1 + i \gamma_2$ is the \textit{gravitational shear}, that quantifies
anisotropic stretching, i.e. distortions of the shape.  

In the context of cosmological lensing by large-scale structures, galaxies are very weakly lensed and the values of $\kappa$ and $\gamma$ are on the order of few percent. The observable of the cosmic shear is based on the measurement of galaxy shapes, i.e. the \textit{reduced shear}:
\begin{equation}
\mathbf{g}=\frac{\mathbf{\gamma}}{1-k},
\end{equation}
which has the same polar (i.e. spin-2) transformation properties as shear.

Galaxies are intrinsically non-circular in general, so an intrinsic, complex \textit{source ellipticity}~$\mathbf{e}^s$ can be attributed to a galaxy. 
If we define ellipticity $e=(a-b)/(a+b)$, for galaxy major axis $a$ and minor axis $b$, then
the observed ellipticity under the gravitational lens mapping is  given by~\citep{SS97}: 
\begin{equation}
\mathbf{e} = \frac{\mathbf{e}^s + \mathbf{g}}{1+\mathbf{g}^*\mathbf{e}^s},
\end{equation}
where both ellipticity and shear are defined as complex numbers encoding the shape in the absolute value and the orientation in the phase, i.e. $\mathbf{e}=e \exp{(2i\theta)}$. 

Under the assumption of randomly oriented galaxies, $\langle \mathbf{e}^s \rangle = 0$, 
the observed ellipticity is an estimator of the gravitational shear:  
$\langle \mathbf{e} \rangle = g \simeq \gamma$, in the weak lensing
regime $|\gamma|,\kappa \ll 1$.
The typical distortion of high-redshift galaxies by the gravitational potential is much smaller than the intrinsic dispersion in galaxy shapes ($\sigma_e = \langle |\mathbf{e}|^2\rangle^{1/2} \sim 0.3$). Thus, for an individual galaxy, the lensing effect is not detectable and one needs to average over a large number of galaxies $N$ to obtain sufficient signal-to-noise ratio $\textrm{SNR}\simeq g \times N^{1/2}/\sigma_e$.

The shear is estimated as a weighted average of the galaxies' ellipticities. 
Statistical weights take into account that faint galaxies have broader likelihood surfaces (i.e.  larger measurement errors) than bright galaxies.   
We calculate an approximate inverse-variance weight as defined in Sect.~3.6 in~\citet{Miller13}:
\begin{equation}\label{shear}
 w_i=\Big[\frac{\sigma_i^2 e^2_\textrm{max}}{e^2_\textrm{max}-2\sigma_i^2} + \sigma^2_\textrm{pop}\Big]^{-1}
\end{equation}
where $e_{\max}$ is the maximum allowed ellipticity (as measured in the same paper for the prior ellipticity distribution), $\sigma^2_i$ is the 1D variance of 
the likelihood for the $i$-th galaxy (\textit{measurement noise} of galaxy $i$) and $\sigma^2_\textrm{pop}$ is the 1D variance of the ellipticity distribution of the observed galaxy population (\textit{shape noise}). We define these 1D variances as the square root of the covariance matrix determinant. Notice that, in the limit where $e_\textrm{max} \rightarrow \infty$, this definition of the weights tends to a conventional form $w_i \rightarrow (\sigma_i^2+\sigma^2_\textrm{pop})^{-1}$.
 
\section{The RadioLensfit method}
\label{sec:radiolensfit}
In the radio implementation of \textit{lens}fit, we apply the method directly in Fourier space, where the radio interferometer data is measured. In particular, we use a galaxy model defined analytically in the visibility domain (Sect.~\ref{sec:model}) and marginalise the likelihood over uninteresting parameters such as flux, galaxy position and galaxy scalelength (Sect.~\ref{sec:fitting}). Finally, we sample the resulting likelihood~(Sect.~\ref{sampling}) as a function of the ellipticity parameters only, in order to estimate the galaxy ellipticity as the likelihood mean point and to compute the likelihood standard deviation as the corresponding measurement noise, $\sigma_i$. This approach is due to the fact that a likelihood estimator of the ellipticity of an individual galaxy should respond linearly to a cosmological shear, whereas a posterior estimation would lead to a bias in the shear measurement, as discussed by \citet{Miller13}.

\subsection{Analytical galaxy model in the visibility domain} 
\label{sec:model} 
As mentioned in the introduction, galaxy models in the radio regime should approximate the optical disk component, which is well described and commonly used for optical weak lensing, by the S\'{e}rsic exponential brightness profile:
\begin{equation}\label{profile}
I(r) = I_0 \exp(-r/\alpha),
\end{equation}
where $I_0$ is the central brightness and $\alpha$ is the scalelength (i.e. the radius at which intensity drops by $e^{-1}$). This function defines a circular light profile, with coordinates $(l_r,m_r)$, that is made elliptical and rotated according to the ellipticity parameter $\mathbf{e} =(e_1,e_2)$ using the following linear transformation:
\begin{equation}\label{linearTransf}
  \left( \begin{array}{c} l_r \\  m_r \end{array} \right) =  A \mathbf{x} = \left( \begin{array}{cc} 1-e_1 & -e_2 \\ -e_2 & 1+e_1 \end{array} \right)  \left( \begin{array}{c} l \\ m \end{array} \right).
\end{equation}
The galaxy image obtained can be sampled by computing the direct Fourier transform at the \textit{uv} points of the radio telescope. 
However, due to the simplicity of this brightness profile, we are able to directly define this model in the Fourier space by computing the analytical expression of the Fourier transform ($\mathcal{F}$) of function (\ref{profile}).
Since it is a circularly symmetric function, its Fourier transform is essentially its Hankel transform of order zero~$\mathcal{H}_{(0)}$:
\begin{align} 
\mathcal{F}(I(r))(k) & = \int_{r=0}^\infty \int_{\theta=0}^{2\pi} I_0 e^{-r/\alpha} e^{-2\pi ikr\cos{\theta}} r dr d\theta \nonumber \\
     & = 2\pi I_0 \int_{r=0}^{\infty} e^{-r/\alpha} J_0(2\pi kr)r dr \nonumber \\ 
     & =  2\pi I_0 \mathcal{H}_{(0)}(e^{-r/\alpha})(2\pi k) 
\end{align}
where $J_0$ is the Bessel function of order zero.
For the exponential function the Hankel transform is well known~\citep{Hankel}:
\begin{equation}
\mathcal{H}_{(0)}(e^{-ar}) (k)= \frac{a}{(a^2 + k^2)^{3/2}}.
\end{equation}
Therefore
\begin{equation}
\mathcal{F}(I(r))(k) = \frac{2\pi \alpha^2 I_0}{(1+4\pi^2 \alpha^2k^2)^{3/2}}.
\end{equation}
Finally, by applying the following result for the composition of a function with a linear transformation defined by a matrix $A \in \mathbb{R}^{d \times d}$:
\begin{equation}
\mathcal{F}(I \circ A)(\mathbf{k}) = \frac1{\det A} \mathcal{F}(I(r))(A^{-T}\mathbf{k}),
\end{equation}
where $A^{-T}$ is the inverse transpose of $A$, we get the following expression of the visibility produced by the observed galaxy at the point $\mathbf{k}=(u,v)$: 
\begin{align}
V(u,v) & = \mathcal{F}(I \circ A)(\mathbf{k}) \nonumber \\
         & = \frac{2\pi \alpha^2 I_0}{\det A \big(1+4\pi^2 \alpha^2|A^{-T}\mathbf{k}|^2\big)^{3/2}}
\end{align}
In terms of flux density at wavelength $\lambda$, we have $S_{\lambda} = 2\pi \alpha^2 I_0/\det A$.
Moreover, if we want to take into account the source spectrum, to first order we can model it using a single spectral index~$\beta$ as follows:
\begin{equation}\label{galaxy}      
 V(u,v) = \Big( \frac{\lambda_\textrm{ref}}{\lambda}\Big)^\beta \frac{S_{\lambda_\textrm{ref}}}{\big(1+4\pi^2 \alpha^2 |A^{-T}\mathbf{k}|^2\big)^{3/2}}.
\end{equation}
where $\beta=-0.7$ for the synchrotron radiation emitted by the galaxy disk at $\sim 1$~GHz. We may reasonably assume the
spectral index to be invariant with frequency across the SKA bandpass, because the intrinsic synchrotron spectrum is broadband and featureless.

\subsection{Bayesian marginalisation of the likelihood}
\label{sec:fitting}
The model visibilities depend on 6 parameters: flux $S$, scalelength $\alpha$, centre position $\mathbf{c} = (l_0,m_0)$ and ellipticity $(e_1,e_2)$.
Since we are interested only in the measurement of the ellipticity, we can marginalise over the other parameters.
To compute the likelihood $\mathcal{L}$ we adopt a chi-squared fitting approach in the frequency domain, where the visibilities are defined:
\begin{equation}
\begin{split}
\chi^2 & =  (D-S M)^{\dagger}C^{-1}(D-S M) \\
         & = D^{\dagger}C^{-1}D - 2S D^{\dagger}C^{-1}M + S^2 M^{\dagger} C^{-1}M 
\end{split}
\end{equation} 
where $D=(v_j)_j$ and $M=(v_j^m)_j$ are respectively the data and flux independent model visibilities, and $C$ is the noise covariance matrix.\footnote{In Appendix~\ref{Matrix-cov} we show that for weak sources, as measured in weak lensing surveys, the visibility noise covariance matrix may be assumed to be diagonal.}
By normalising the model visibilities by a factor $(M^{\dagger}C^{-1}M)^{-1/2}$, we can write chi-squared as
\begin{equation}
\begin{split}
\chi^2  = & D^{\dagger}C^{-1}D + \Big[S - \frac{D^{\dagger} C^{-1}M}{(M^{\dagger}C^{-1}M)^{1/2}}\Big]^2 \\
& - (M^{\dagger} C^{-1}M)^{-1}(D^{\dagger}C^{-1}M)^2. 
\end{split}
\end{equation}
We marginalise the corresponding likelihood over $S$ by assuming a uniform prior for the flux. Since the function to be integrated is Gaussian-like and we expect to be measuring radio sources that have a significant ($> 10\sigma$) 
detection of their
radio flux, we may integrate over the range $(-\infty, \infty)$ for which the result of the integration is well-known. Therefore we have
$$
\mathcal{L} = e^{-\chi^2/2} \propto \exp \Big[\frac12 (M^{\dagger} C^{-1}M)^{-1}(D^{\dagger}C^{-1}M)^2\Big].
$$
The shift parameter ${\mathbf x}$ of the model position can be added in the Fourier domain by multiplying the model visibilities by a factor $e^{i\mathbf{k}^T\mathbf{x}}$. Therefore 
\begin{equation}
\label{cross-corr}
 \mathcal{L} \propto \exp \Big[\frac12 (M^{\dagger} C^{-1}M)^{-1}h(\mathbf{x})^2\Big],
 \end{equation}
where $h({\mathbf x}) =D^{\dagger}C^{-1}(v^m_j e^{i\mathbf{k_j}^T \mathbf{x}})_j$
corresponds to the Fourier transform of the \textit {cross-correlation} function in the image domain.
As discussed by \citet{Miller07},
such a cross-correlation should be well represented by a 2-dimensional Gaussian function, therefore by evaluating its maximum $h_0=h(l_0,m_0)$ we can approximate $h({\mathbf x})$ as a real analytical function. 
\begin{equation}
h(\mathbf{x})\sim  h_0\exp{\Big[-\frac12 (\mathbf{x}-\mathbf{c})^T \Sigma^{-1}(\mathbf{x}-\mathbf{c}) \Big]}
\end{equation}
and use it to analytically marginalise the likelihood over the position shift parameter assuming a uniform prior.
\begin{equation}
\begin{split}
\log \mathcal{L} = k\exp{\Big[- (\mathbf{x}-\mathbf{c})^T \Sigma^{-1}(\mathbf{x}-\mathbf{c})\Big] } + \textrm{const}, \\
 k = \frac{h_0^2}2 (M^{\dagger} C^{-1}M)^{-1}.  
\end{split}
\end{equation}  
Then by using polar coordinates and a uniform prior $P(r)$ over the area $\pi r_\textrm{max}^2$, we have
\begin{equation}
\label{marg-shift}
 \int_0^{2\pi} d\theta \int_0^{r_\textrm{max}}\mathcal{L}(r) P(r)rdr \propto \frac{|\Sigma|^{1/2}}{r_\textrm{max}^2} \int_0^{r_\textrm{max} } \exp \left[ ke^{-r^2}\right] rdr.
\end{equation}
Following \citet{Miller13},
the maximum position uncertainty $r_{max}$ over which to marginalise is chosen to be the position beyond which the detection of the galaxy becomes statistically insignificant, e.g. corresponding to the 95\% confidence region for the location of the galaxy, according to the likelihood-ratio test:
\begin{equation}
-2\log \left( \frac{\mathcal{L}(r_\textrm{max})}{ \mathcal{L}(0)}\right)  = \chi_\textrm{crit}^2 \quad \text{with 2 d.o.f.}
\end{equation}
i.e.
$$
-2k (e^{-r_\textrm{max}^2} - 1) = \chi_\textrm{crit}^2 = 5.991.
$$
Equation~(\ref{marg-shift}) is solved, after the substitution $t=-ke^{-r^2}$, as an exponential integral that can be evaluated numerically.
The cross-correlation maximum point $h_0=h(l_0,m_0)$ is computed using the Newton method and $|\Sigma|$ is obtained as the inverse of the determinant of the Hessian matrix of $h({\mathbf x})$ at the maximum point.
Finally, a marginalisation over a finite interval $[\alpha_\textrm{min}, \alpha_\textrm{max}]$ of the scalelength is computed numerically assuming a lognormal prior (see Sect.~\ref{data-sim}).

\subsection{Likelihood sampling}
\label{sampling}
In order to measure the ellipticity and its uncertainty for each galaxy, we measure the likelihood standard deviation by sampling a neighbourhood of its maximum point.

The likelihood maximum point is computed by applying the simplex method~\citep{Nelder-Mead}. We use the implementation of this algorithm provided by the GNU Scientific Library\footnote{http://www.gnu.org/software/gsl/} with tolerance $\textrm{tol} = 10^{-3}$ and starting point $(0,0)$.
We adopt an adaptive grid sampling: the likelihood is first estimated on a coarse grid of ellipticity values with step~0.05, then the grid step is iteratively reduced by a factor~2, until either at least~30 points had been measured above a threshold of 5\% of the maximum likelihood or a resolution of 0.003 in ellipticity is reached. The likelihood threshold is also used to define the size of the neighbourhood where sampling. 
As an estimate of the galaxy's ellipticity, we compute the mean of the likelihood distribution using the ellipticity samples above the threshold, and for each ellipticity component we take as the measurement error the 1D standard deviation of the likelihood.   

\section{Data simulations}
\label{sec:ska-simulations}
\subsection{SKA1-MID specifications}
\label{ska1-mid}
\begin{figure}
\centering
\includegraphics[scale=0.45]{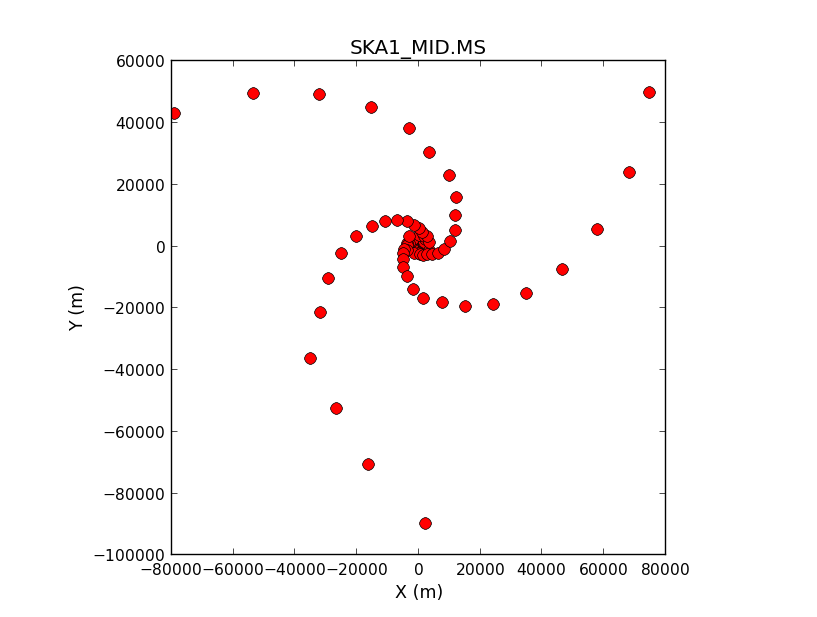}
\caption{SKA1-MID antennas location.}
\label{fig:antennas}
\end{figure}

\begin{figure}
\centering
\includegraphics[scale=0.45]{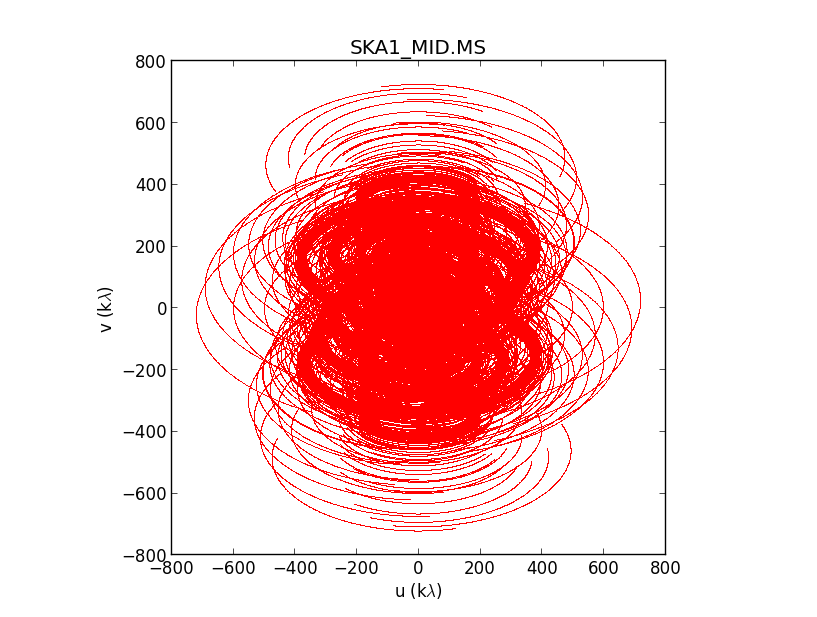}
\caption{SKA1-MID uv coverage at declination $\delta = -30^\circ$.}
\label{fig:coverage}
\end{figure}

In Phase 1, SKA will consist of two sub-arrays: SKA1-LOW will be an aperture array located in Australia operating at low radio frequencies, while SKA1-MID will be a dish array located in South Africa with up to five observational frequency bands spanning the range 350~MHz to 13.8~GHz \citep{SKA1b}.
For weak lensing surveys, SKA1-MID will be used, as it provides both the sensitivity and the spatial resolution to detect shapes on high redshift SF galaxies. It will comprise 64 MeerKAT dishes in a moderately compact core with a diameter of about 1~km and 133 SKA1 dishes distributed in the core and in three logarithmically spaced spiral arms emanating from the centre and extending out to a maximum radius of 80~km (see Fig.~\ref{fig:antennas}), with a maximum baseline of 150~km. We simulate an 8-hour observation at declination $\delta = -30^\circ$ assuming, for simplicity,  that all the antennas are SKA dishes (so that all the visibilities have the same noise variance) and adopting a natural weighting scheme. We use the first 30\% of Band 2, i.e. 950 - 1190 MHz, as proposed in~\citet{BHCB16}, and sample visibilities every $\tau_\textrm{acc} = 60$~s for 12 channels of bandwidth 20~MHz. Figure~\ref{fig:coverage} shows a plot of the uv coverage for our simulations. 
 
\subsection{Simulated galaxy visibilities}
\label{data-sim}
Since galaxies cannot be individually distinguished in the visibility domain, in this first paper we only simulate visibilities of individual galaxies located at the phase centre, to test the effectiveness of the method in this simplest case.
Future developments will need to fit to multiple galaxies observed within the primary beam 
(Sect.~\ref{sec:multiple}).

The flux and scalelength of the simulated galaxies are generated randomly according to the distributions estimated in~\citet{Rivi15} from the VLA 20~cm continuum radio source catalog in the SWIRE field. 
The modulus $e$ of the intrinsic ellipticities are generated according to a distribution estimated from 66,762 SDSS disk-dominated galaxies (see Appendix B2 in \citet{Miller13}):
\begin{equation}
P(e) = \frac{Ne\left(1-\exp\left[ \frac{e-e_\textrm{max}}{c}\right]\right)}{(1+e)(e^2+e_0^2)^{1/2}},
\end{equation}
where $e_\textrm{max}=0.804$, $e_0=0.0256$, $c=0.2539$ and $N$ is a normalisation factor. 
Optically-selected galaxies are subject to the effect of luminosity-dependent,
inclination-dependent obscuration, that suppresses the prevalence
of high-ellipticity galaxies in optical surveys. Thus
we expect the ellipticity distribution of radio galaxies to be different from the optical regime, and likely to 
extend to higher ellipticities,
but at the moment we have no information about the ellipticity distribution of the faintest radio-selected galaxies,
and for simplicity we use the distributions assumed for galaxies in optical surveys. Our conclusions on the utility
of the method should not be dependent on this choice.
For each ellipticity modulus value, 
10 equally-spaced galaxy orientations are defined, starting from a random angle value $\theta_0 \in (0,\pi)$ generated according to a uniform distribution, so that the corresponding ellipticity values are distributed uniformly on a circle. 
Keeping the same size and flux of the galaxies whose ellipticity values are on the same ring, 
the simulations are largely free of shape noise, i.e. the unweighted average of the intrinsic ellipticity is identically zero,
$\langle\mathbf{e}^s\rangle = 0$, 
and the unweighted average of the sheared ellipticity yields the input shear to a good approximation:
$\langle\mathbf{e}\rangle \simeq \mathbf{g}$.\footnote{Actually the effectiveness of this shape noise cancellation is reduced by the SNR dependence on galaxy orientation (because we do not sample the Fourier modes isotropically), which does affect the galaxy weights that are used in the shear computation.} Reduction of shape noise significantly reduces the volume of simulations required to evaluate the shear measurement accuracy.

Visibilities of real observations are simulated by using equation~(\ref{galaxy}) as for the model, but  adding an uncorrelated gaussian noise whose variance is dependent on the SEFD of the SKA antennas (see~Appendix~\ref{Matrix-cov}). No time or frequency smearing effects are included, as the galaxies are assumed to be located at the phase centre, where such smearing is negligible.

\subsection{Gridding visibilities}
Since the number of visibilities are very large (more than $10^4$ per time sample per frequency channel), directly using all the uv data is very expensive both in terms of memory size and computational time.
For this reason we apply a gridding scheme to reduce the data volume.
We have defined a regular uv grid of size $n\times n$ and taken the average of the $c_i$ observed visibilities falling in the same grid cell $i$. 
This operation reduces the variance (if assumed the same for every uv point) of each grid visibility $\bar v_i = \langle v_{k_i} \rangle_{c_i}$ by a factor $c_i$ and therefore equation~(\ref{cross-corr}) becomes:
\begin{equation}
 \mathcal{L} \propto \exp \left[\frac{\left[  \sum_i \Re (\bar v_i^* v_i^m e^{-ik {\mathbf x}})c_i\right]^2}{2\sigma_v^2\sum_i |v_i^m|^2 c_i}\right].
\end{equation}
Model visibilities are sampled on the gridded uv points and only non zero visibilities of the grid are considered. 
Usually the cell size is chosen to be $\Delta u = \Delta v = 1/\psi$, where $\psi$ is the intended field-of-view of one beam at the band centre. This choice would minimise the number of cells avoiding smearing at large scales, which mimic primary beam attenuation. However for our specific case (one galaxy at the phase centre) we can consider a coarser grid. 
By testing the shape fitting for the same visibilities with different grid sizes, we obtain $n=800$ as the smallest size we can use~(see \citealt{Rivi15}).  

\subsection{Code implementation}
The C++/C code implemented for simulating visibility data and fitting galaxy shapes is available online\footnote{http://github.com/marziarivi/RadioLensfit}. It has been parallelized in a hierarchical way by using the Message Passing Interface\footnote{http://www.mpi-forum.org/} (MPI) and OpenMP\footnote{http://openmp.org/} parallel programming paradigms, enabling the user to exploit HPC architectures. The first level of parallelisation (MPI) simply distributes the simulated galaxies among different nodes/multi-core processors, each simulating and performing the model fitting of its own chunk of galaxies. In the second level (OpenMP), each thread computes visibilities for a different channel, which is the most computing intensive part of the code.
The main reason for such hybrid implementation is to exploit all the CPU cores used when a large amount of RAM is required. This happens in the realistic case where galaxies are located randomly in the field of view because visibilities must be gridded in a very large grid and the uv sampling before gridding must be higher, to minimize time and frequency smearing effects.

\section{Results}
\label{sec:shear}
\begin{figure}
\centering
\includegraphics[scale=0.20]{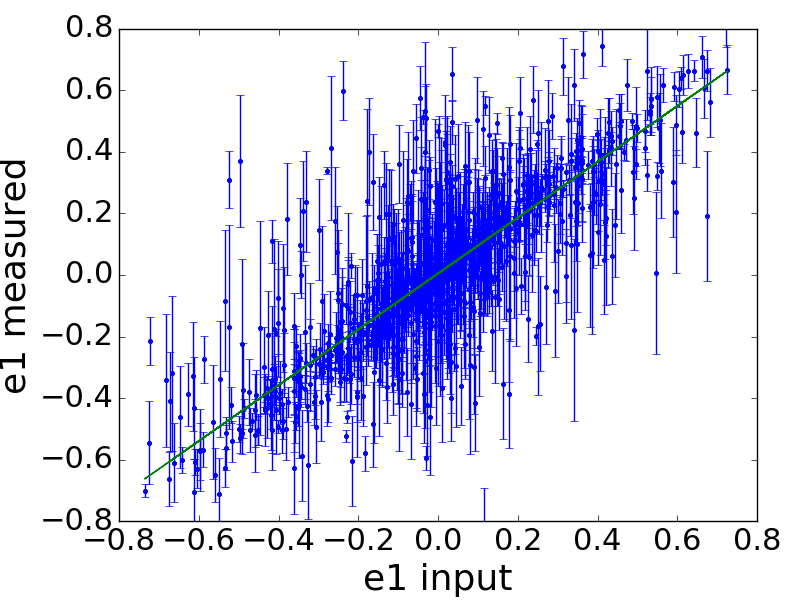}
\includegraphics[scale=0.20]{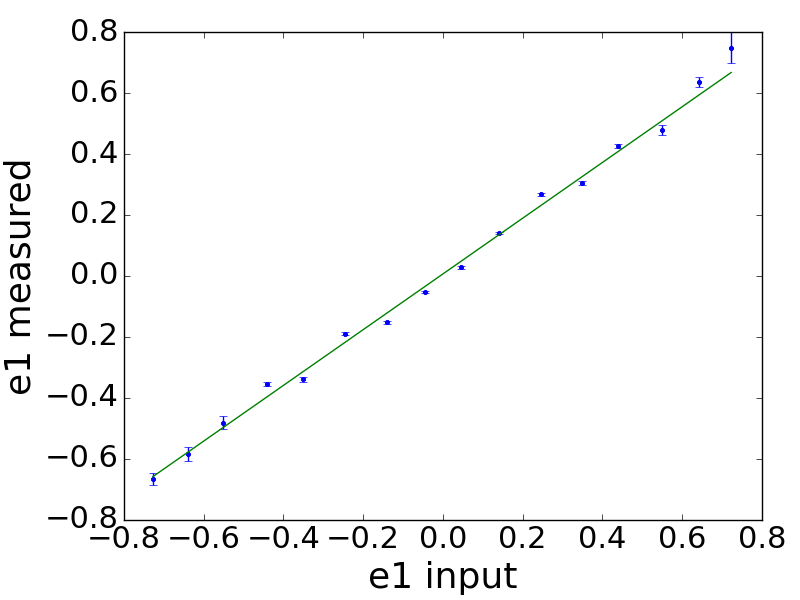}
\includegraphics[scale=0.20]{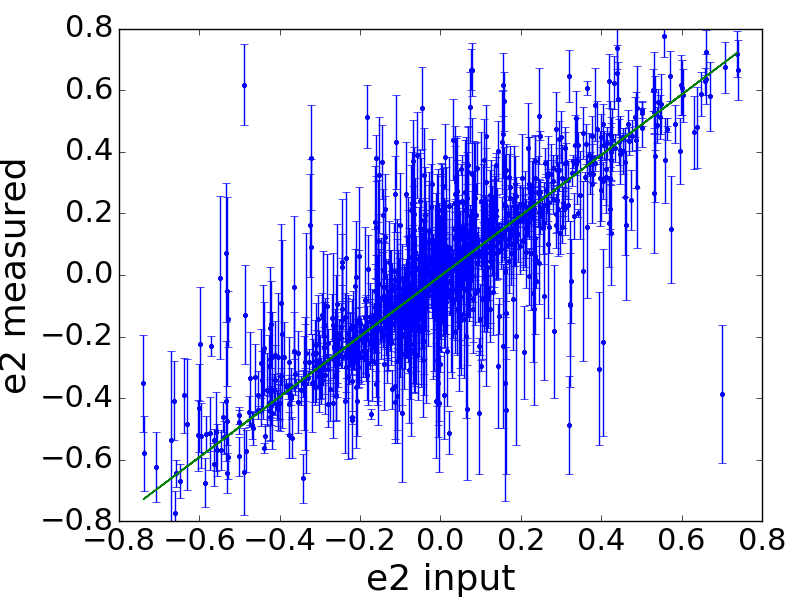}
\includegraphics[scale=0.20]{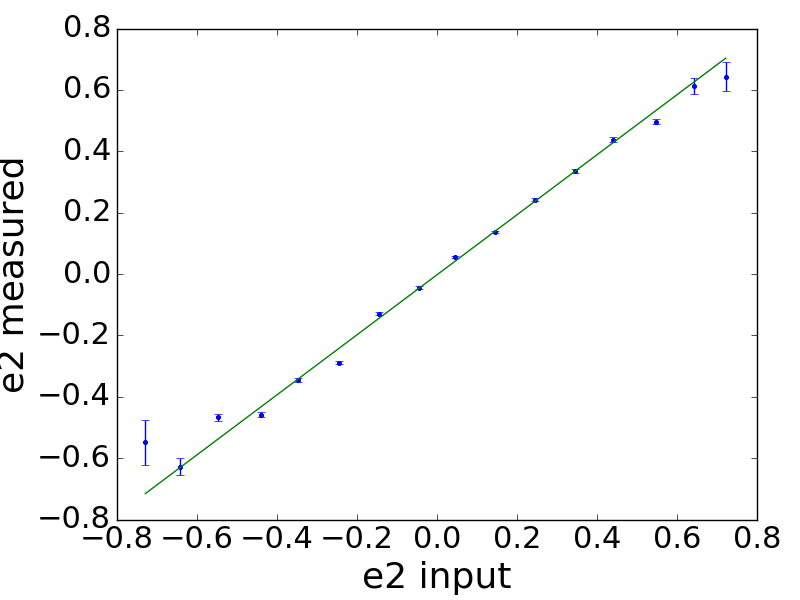}
\caption{Plots for both components of 1000 galaxy shapes fitting input sources with flux density $S\ge 10 \mu$Jy. The left side shows the likelihood mean and standard deviation of the measured ellipticities plotted against the input values. The right sides show the same values binned.}
\label{fig:shapesS10}
\end{figure}
\begin{figure}
\centering
\includegraphics[scale=0.20]{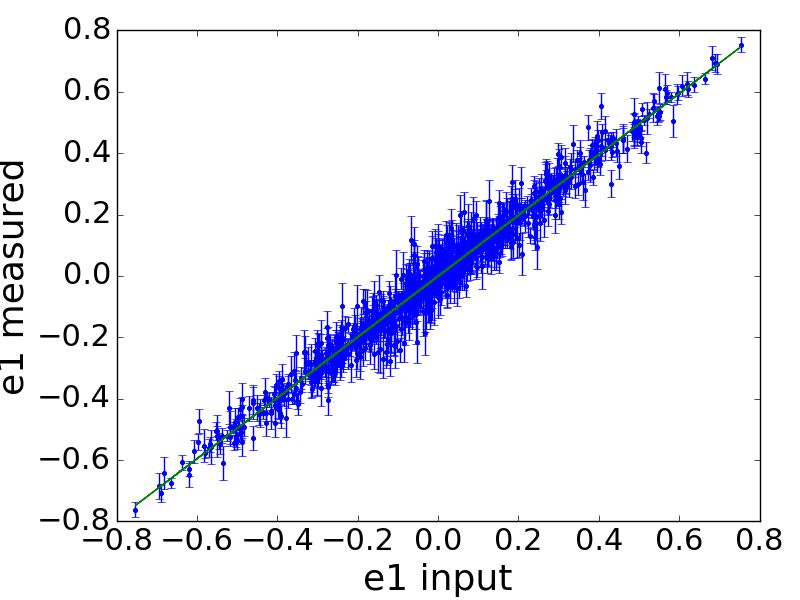}
\includegraphics[scale=0.20]{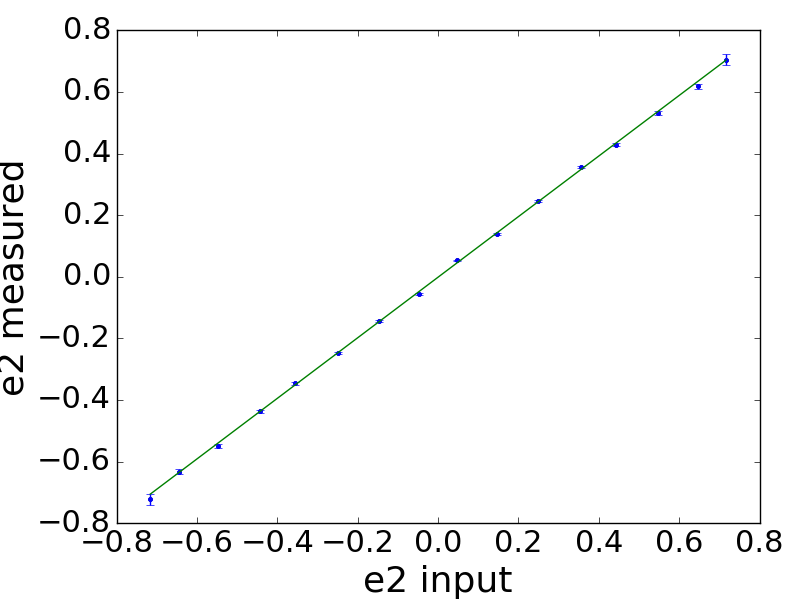}
\caption{Plots for the $e_1$  component of galaxy shapes fitting input sources with flux $S > 50 \mu$Jy. The right sides show the corresponding binned values. Results for $e_2$ component are similar.}
\label{fig:shapesS50}
\end{figure}
We simulate populations of individual galaxies whose flux densities lie in the interval $10-200\mu$Jy, corresponding to SNR\footnote{We compute the signal-to-noise ratio in the visibility domain as $\textrm{SNR} = \sqrt{\sum_{i=1}^\textrm{nvis} |v_i|^2/\sigma^2_i}$, where $v_i$ are the visibilities without noise.}\,$\ge 10$, and compare input and measured galaxy ellipticity values. 
This flux range has been chosen because of the telescope sensitivity (lower bound) and in order to simulate faint galaxies with redshift $z > 0.5$ (upper bound), the most relevant ones for radio weak lensing~\citep{BHCB16}.

\begin{figure}
\centering
\includegraphics[scale=0.43]{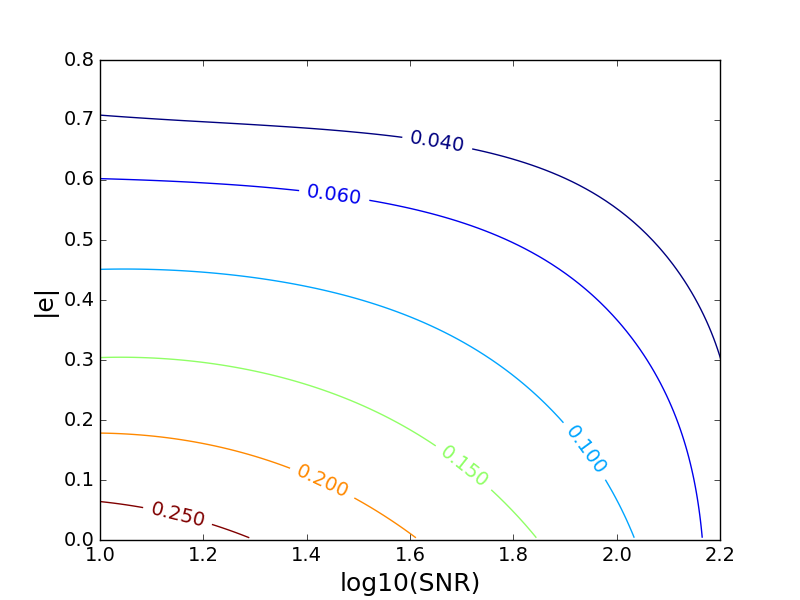}
\caption{Distribution of the measurement 1D standard deviation showing the dependence on the source ellipticity and signal-to-noise ratio.}
\label{fig:errors}
\end{figure}

Fig.~\ref{fig:shapesS10} shows measurements of both ellipticity components for each simulated galaxy. The slopes of the best-fit lines are respectively $0.9065 \pm 0.0060$ and $0.9807 \pm 0.0059$.
As expected, at higher signal-to-noise ratios (see Fig.~\ref{fig:shapesS50} for a population with galaxy flux $S > 50 \mu$Jy) there is a better correspondence between input and output values. The best-fit slopes, respectively $0.9914 \pm 0.0035$ and $0.9864 \pm 0.0035$, are closer to unity and measurements have a reduced dispersion.
Fig.~\ref{fig:errors} shows a contour plot of the measurement noise of $10^4$ galaxy shapes highlighting a strong dependence on the source SNR and ellipticity value. As expected, accurate model fits are more difficult for round galaxies at low SNR.

\begin{figure}
\centering
\includegraphics[scale=0.38]{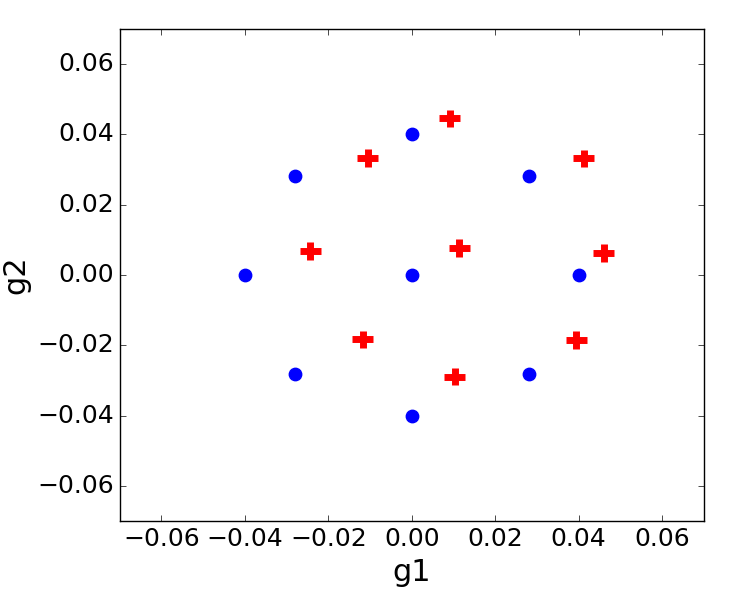}
\includegraphics[scale=0.38]{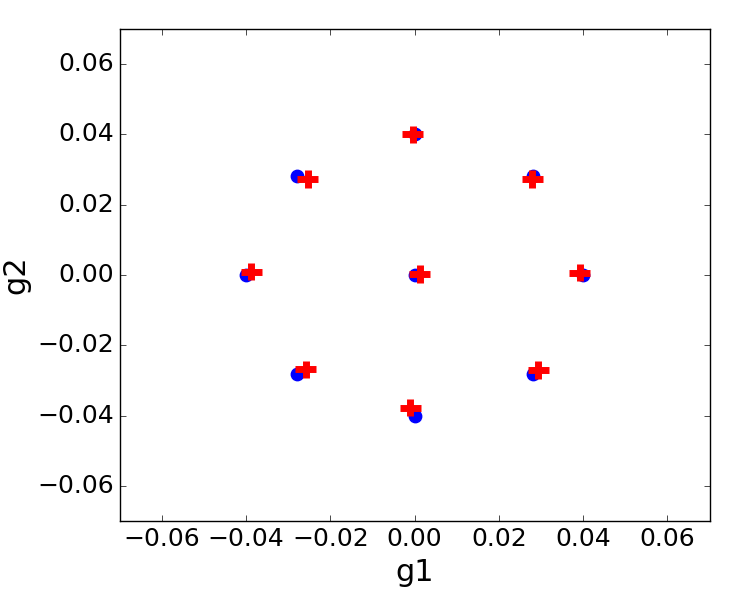}
\caption{Shear measurements from which the noise bias is computed; input values are blue points while measured values are red crosses.
 {\em Top}: $\textrm{SNR}>10$. {\em Bottom}: $\textrm{SNR}>25$.} 
\label{fig:shear}
\end{figure}

To measure the reduced shear~$\mathbf{g}$ with a statistical uncertainty of 1\%, we generate populations of $10^4$ galaxies.
Error bars correspond to the standard deviation of the shear values estimated from 1000~bootstrap resamples. 
Typically, 6\% of galaxies are not measured as their likelihoods are too noisy: these are given zero weight in the
analysis. We estimate the shear measurement bias for this method by applying an input reduced shear with amplitude $g=0.04$, so that we should be in a linear regime of bias measurement. We consider 8 different shear orientations and also the case $g=0$, as plotted in Fig.~\ref{fig:shear}. 
Input and measured shear ellipticity values are compared assuming a linear bias model
\citep{Heymans06},
\begin{equation}
g_i^m - g_i = m_i g_i + c_i, \qquad i=1,2,
\end{equation}
where $g_i^m$ (resp. $g_i$) is the $i$-component of the measured (resp. original) value of the input reduced shear, $m_i$ and $c_i$ are respectively the multiplicative and additive biases.
A non-zero multiplicative bias indicates calibration errors due to 
effects such as noise bias (\citealt{Melchior12}, \citealt{Refregier12}) or weight bias
(Fenech Conti et al., in preparation) and a non-zero additive bias indicates a systematic error 
due to effects such as the correlation of noise bias with the PSF \citep{Miller13}. 

Shear bias estimates obtained from the best-fit lines of shape measurements at $\textrm{SNR}>10$ are:
\begin{align}
& m_1 = 0.101  \pm 0.018, \quad c_1 = 0.0123  \pm 0.0005; \nonumber \\ 
& m_2 = 0.080 \pm 0.018, \quad c_2 = 0.0073  \pm 0.0005. \nonumber
\end{align} 
These results show different bias values for the two ellipticity components, probably due to the asymmetry of the uv coverage (see Fig.~\ref{fig:coverage}), confirming a better accuracy on the second component at low SNR, as for the galaxy shapes. 
The top panel of Fig.~\ref{fig:shear} shows the corresponding plots of the shear measurements compared with the input values. 
Similarly, we compute the shear bias from simulated populations with the same parameters distributions but with different lower limit fluxes, in order to investigate its relation with minimum SNR. The lower panel of Fig.~\ref{fig:shear} shows how the shear measurements improve for $\textrm{SNR}>25$, where the measured shear bias is: 
\begin{align}
& m_1 = 0.0281  \pm 0.0098, \quad c_1 = 0.00065  \pm 0.00026; \nonumber \\ 
& m_2 = 0.0318 \pm 0.0098, \quad c_2 = 0.00054  \pm 0.00026. \nonumber
\end{align}

In \citet{Brown15}, requirements for the shear multiplicative and additive bias for SKA and other future optical surveys are estimated  by considering three parameters: sky area, galaxy median redshift and galaxy number density. These requirements are set such that cosmological results are dominated by statistical rather than systematic errors and therefore they define an upper limit on the level of bias accuracy. The sensitivity levels have been chosen appropriately for image domain resolution of 0.5\,arcsec at Band~2 and the galaxy number densities correspond to $>10\sigma$ detections.
For a 2-year continuum survey with SKA1-MID over $5000~\deg^2$ (as proposed for weak lensing in~\citealt{SKA1b}) and $z_\textrm{med}=1.0$, the following constraints are obtained: multiplicative bias $m < 0.0067$, additive bias $c < 0.00082$.

\begin{figure}
\centering
\includegraphics[scale=0.45]{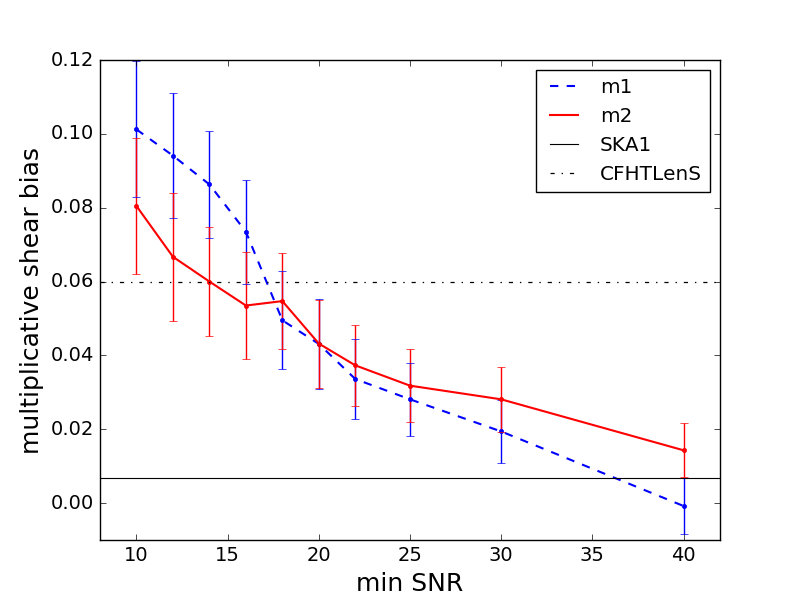}
\caption{Plot for both components of the multiplicative shear bias for different minimum SNR values. They are compared with SKA1 bias requirements (continuum black line) and CFHTLenS calibration correction (dash dot line).} 
\label{fig:multiplicative-bias}
\end{figure}

\begin{figure}
\centering
\includegraphics[scale=0.45]{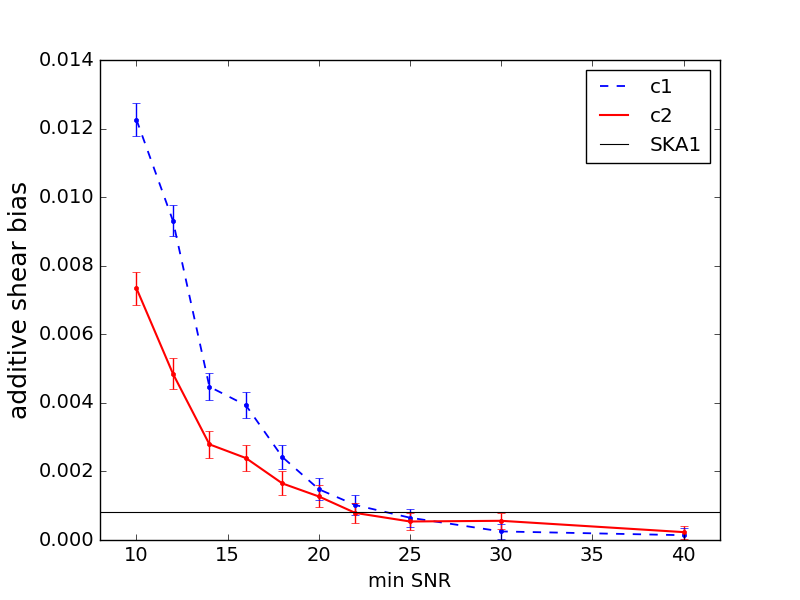}
\caption{Plot for both components of the additive shear bias for different minimum SNR values.} 
\label{fig:additive-bias}
\end{figure}

A plot of the multiplicative and additive shear biases are shown respectively in Fig.~\ref{fig:multiplicative-bias}  and Fig.~\ref{fig:additive-bias}. They show how the measured bias components decrease as the SNR lower limit increases up to~40. The additive bias turns to be comparable with SKA1 requirements for $\textrm{SNR}>18$, while the multiplicative bias starts to be comparable with SKA1 requirements for $\textrm{SNR}>30$.  

At the lowest SNR values, the current method displays a multiplicative bias as expected from noise bias \citep{Melchior12, Refregier12}.
Existing optical weak lensing surveys have biases on shear exceeding their cosmology requirements, primarily as a result of noise bias, for which an accurate mathematical correction has not yet been devised, except possibly by averaging over many galaxies \citep{BA2014}. In those surveys, the approach that has been taken is to derive a calibration for the noise bias from simulations and apply that calibration to the measurements of the data.
We may compare our results with the correction for shear measurement bias that was made for CFHTLenS \citep{Heymans12}. This is a ground-based optical survey with lensing data in the optical $i$-band for galaxies with $\textrm{SNR}\ge 10$. Shear was measured by a model-fitting method using {\em lens}fit with a multiplicative bias correction that was dependent on both galaxy signal-to-noise ratio and size. The weighted average multiplicative bias correction was $m \simeq 0.06$, which is comparable to our results.  We expect that a post-measurement shear calibration correction may then
be applied to the radio waveband measurements that is comparable to that needed for the current generation of optical lensing surveys.

The additive bias found in the radio waveband measurements is already close to requirements, but we note that in these simulations, an additive bias can only arise from the anisotropic sampling of the visibility plane. This effectively results in an
anisotropic image domain PSF, and we expect an additive noise bias component
in this case \citep{Miller13}.  However, in radio interferometer data this effect may
be eliminated or mitigated by weighting the samples in the visibility domain to improve
the isotropy of the measurement: an advantage of interferometer measurements that is
not possible with optical lensing surveys. Additive biases are particularly important
on large angular scales in the cosmic shear correlation function, and currently
residual additive biases limit the maximum angular scales that may be probed in
cosmological analyses \citep{H13, Kilbinger13}. In the case of SKA, the visibility 
domain anisotropy will be a function of the declination of the field being observed
(here we have simulated a field that transits close to the zenith),
so larger isotropy weighting corrections may be needed at more extreme observation
declinations.

\section{Discussion on the fitting of many sources in the visibility domain}
\label{sec:multiple}

A full analysis method should be able to measure the shapes of all galaxies in the field of view. 
The main challenge for SKA is the large number of sources contained in the primary beam (up to $10^4$ for SKA1). 
We envisage an initial imaging step would allow identification of the locations of all detected sources in the field, and those
positions could then be used to fit models to the visibility data. 
For fitting the shapes of so many sources we propose two possible approaches: either joint fitting  all of them, taking into account the computational effort, or following the optical case by extracting a single or a small group of clustered galaxies at a time, which are much easier to fit simultaneously, taking into account possible effects that such a procedure could introduce in the data. We plan to investigate both directions as follows.

\subsection{Joint fitting of all sources}
A Bayesian method able to deal with a large number of parameters is Hamiltonian Monte Carlo (HMC). It exploits techniques developed for Hamiltonian dynamics to suppress random walk behaviour in the distribution sampling and maintain a reasonable efficiency even for high dimensional problems. For a review about the method see~\citet{Neal11}.
HMC has been already used in the estimation of the CMB power spectrum from simulated WMAP data, where it has been able to fit $2\cdot 10^5$ parameters and performed favourably even at low signal-to-noise \citep{TAH08}.
This method requires the gradient of the distribution that is sampled, that in our case will be the likelihood function. We can provide it analytically because the visibility model is the sum, over the number of sources $N$, of the visibilities at the phase centre $V_0$ (given by eq.~(\ref{galaxy})), phase shifted at the position coordinates $(l_s,m_s)$ of each source: 
\begin{equation}
V(u,v)=\sum_{s=1}^N V_0(u,v)e^{-\frac{2\pi i}{\lambda}(u l_s + v m_s)}, 
\end{equation}
where the parameters are the ellipticity components, scalelength and (possibly) the spectral index of the sources.

\subsection{Extraction of a single or few sources}
For the other approach we propose to select each source by following the \textit{faceting} technique \citep{CP92}, already established for the SKA imaging pipeline in order to make image computation feasible and reduce the wide-field problem. It splits up the field of view into a number of facets by phase shifting the visibilities, so that the new pointing direction is at position of interest, and gridding them in a coarse grid (whose size is dependent on the size of the facet). In this way, the contribution to each visibility from sources far from the new phase centre is strongly down-weighted by the sinc Fourier transform of the tophat gridding function.
In this approach, we would select a source by shifting the phase centre and using a grid size similar to the one adopted in this work
(this procedure is analagous to the optical survey approach of extracting a postage stamp for each galaxy in the image domain and Fourier Transforming it to the visibility domain). 

A limitation of this approach might be contamination within a faceted region from bright sources in the field, with sidelobes
passing through the region.
It may be possible to remove such contamination by first {\sc clean}ing the large-scale image, produced as part of the normal
SKA data analysis, before the postage stamp extraction and subtracting the {\sc clean}ed visibilities before extracting
the faceted postage stamp data for the lensing measurement. 

At higher galaxy densities, where there may be multiple galaxies within a faceted region, we may jointly fit a relatively small number of galaxies within each facet, which is more tractable than joint fitting to thousands of galaxies. This should be possible at least for the surface density of galaxies in SKA1, where there is no confusion: in fact the synthesised beam FWHM is $\sim 0.5$~arcsec and therefore we expect to have $10^{-5}$ galaxies per beam area. 

We propose that this approach is used to reduce the number of sources to be simultaneously analysed when full joint fitting is not computationally feasible or is too expensive.
 
\section{Conclusions}
We have presented an adaptation to radio observations of {\em lens}fit~\citep{Miller13}, an algorithm for shear measurement for optical weak lensing surveys. Our version of the method, called RadioLensfit, works  directly in the visibility domain, where radio interferometer data are observed, and fits a galaxy model computed analytically as the Fourier transform of a S\'{e}rsic exponential brightness profile. 

We tested this method for the simple case of individual galaxy visibilities using the SKA1-MID baseline configuration for a continuum survey using the first 30\% of frequency Band~2. Simulated galaxies have been located at the phase centre with flux and scalelength values generated randomly according to distributions that we have estimated from the VLA SWIRE catalog. As we have no information about ellipticity distributions of faint galaxies in the radio regime, we have adopted the optical distribution of galaxy ellipticity modulus, while orientations were generated uniformly around circles in the ellipticity plane in order to be free of shape noise.  
We have measured the sensitivity to shear and estimated the noise bias for the RadioLensfit method at various SNR lower limits.  

This work demonstrates that galaxy shape measurement in the visibility domain provides acceptably accurate values, with multiplicative shear bias on average comparable with the calibration correction applied in the ground-based optical survey CFHTLenS, and additive bias comparable to the requirements on a 5000 $\deg^2$ SKA1 survey for $\textrm{SNR}>18$. 
We have noted that additive biases may be better controlled in radio interferometer data than in optical surveys, as
any anisotropy of the visibility data may be mitigated in the shape measurement process by suitable weighting of 
the measurements.

We have discussed possible approaches for the fitting of many sources in the primary beam, proposing either to use HMC or to select a single source, or few clustered sources, at a time with a ``phase shift and gridding'' faceting technique. These approaches will be both investigated in future work.

We also aim to test this method for simulations where individual galaxies are located randomly in the field of view, rather than being at the phase centre, to put constraints on the number of output channels and the visibility grid size, and further test whether such an approach can in principle meet the requirements of a radio weak lensing survey. The results presented here are encouraging in that respect.

\section*{Acknowledgements}
MR acknowledges the support of the Science \& Technology Facilities Council via a SKA grant. FBA acknowledges the support of the Royal Society via a Royal Society URF award. LM acknowledges STFC grant ST/N000919/1.

\bibliography{master}{}
\bibliographystyle{mn2e_trunc8}
 
\appendix
\section{Covariance matrix of the visibilities} 
\label{Matrix-cov}
Following~\citet{RadioAstronomyII}, we can obtain the covariance matrix of the visibility noise by computing the covariance values in a similar way adopted for variances. 
The covariance between the output from two baselines is given by
\begin{equation}
\textrm{cov}(P_{ij},P_{hk}) = \langle P_{ij} P_{hk} \rangle - \langle P_{ij} \rangle \langle P_{hk} \rangle,
\end{equation}
where $P_{ij}$ and $P_{hk}$ are the power for an interferometer involving antennas pairs ($i$,$j$) and ($h$,$k$) after the cross multiplication in the correlator.

If $ij = hk$, then we have the variance for the single baseline~\citep{RadioAstronomyII}:
\begin{align} \label{variance}
\sigma^2(P_{ij}) = & \frac1{2\eta_s^2\Delta\nu \tau_\textrm{acc}} (S_c^2 + S_iS_j + S_i \textrm{SEFD}_j + S_j \textrm{SEFD}_i \nonumber \\
& + \textrm{SEFD}_i \textrm{SEFD}_j),
\end{align}
where $S_i$ is the source flux measured by antenna $i$, $S_c$ is the correlated flux, and SEFD is the antenna {\it System Equivalent Flux Density}, defined as the flux density of a source that would deliver the same amount of power in Jansky of the antenna temperature.

Similarly, if two baselines share one single antenna then
 \begin{equation} \label{covariance}
 \textrm{cov}(P_{ij},P_{jk}) = \frac1{2\eta_s^2\Delta\nu \tau_\textrm{acc}} (S_c^2 + S_cS_j + S_c \textrm{SEFD}_j).
 \end{equation} 
Finally, if the two baselines have no antennas in common then the only contribution to the covariance between their outputs is from the source flux density:
 \begin{equation}
 \textrm{cov}(P_{ij},P_{hk}) = \frac1{\eta_s^2\Delta\nu \tau_\textrm{acc}} S_c^2.
 \end{equation} 
In the weak source regime, $S  \ll \textrm{SEFD}$ and the ratio of a visibility variance to the covariance between such a visibility and another one from a baseline sharing one antenna is approximately $S/\textrm{SEFD}$. For SKA1-MID dishes, whose $\textrm{SEFD} = 400$~Jy, this ratio is of the order of $10^{-3}$ for an amount of $10^4$ faint sources per beam with on average a flux density $S \sim 50\mu$Jy (as observed in weak lensing surveys) and therefore it is negligible.
Therefore for weak lensing observations we can assume the covariance matrix of the visibilities to be  diagonal with
 \begin{equation}
 \label{final_variance}
 \sigma^2(P_{ij}) = \frac1{2\eta_s^2\Delta\nu \tau_\textrm{acc}} (\textrm{SEFD}_i \textrm{SEFD}_j).
 \end{equation}

\end{document}